\documentstyle [12pt] {article}

\catcode`@=12
\begin{document}
\thispagestyle{empty}
\begin{flushright} UCRHEP-T286\\August 2000\
\end{flushright}
\vspace{0.5in}
\begin{center}
{\Large \bf Neutrino Masses and Leptogenesis\\ from $R$ Parity Violation\\}
\vspace{1.3in}
{\bf Ernest Ma\\}
\vspace{0.2in}
{\sl Department of Physics\\ University of California\\ Riverside, CA 92521, 
USA\\}
\vspace{1.3in}
\end{center}
\begin{abstract}\
In $R$ parity violating supersymmetry (conserving baryon number $B$ 
but violating lepton number $L$), Majorana neutrino masses may arise at tree 
level, in one loop, and in two loops.  The $L$ violating interactions work 
together with the $B + L$ violating electroweak sphalerons to erase any 
preexisting $B$ or $L$ asymmetry of the Universe.  To have successful 
leptogenesis nevertheless, a specific scenario is proposed.~\cite{hms}
\end{abstract}
\vspace{0.3in}
\noindent ------------------------

\noindent Talk at 3rd International Conference on Dark Matter in Astro and 
Particle Physics, Heidelberg, Germany (July 2000).

\newpage
\baselineskip 24pt

\section{Introduction}

In the minimal Standard Model, leptons appear under $SU(3)_C \times SU(2)_L 
\times U(1)_Y$ as left-handed doublets $(\nu_i, l_i)_L \sim (1,2,-1/2)$ and 
right-handed singlets $l_{iR} \sim (1,1,-1)$, but there is no $\nu_{iR} \sim 
(1,1,0)$.  Hence any $m_\nu \neq 0$ must necessarily come from the effective 
operator \cite{wein}
\begin{equation}
{1 \over \Lambda} (\nu_i \phi^0 - l_i \phi^+)(\nu_j \phi^0 - l_j \phi^+),
\end{equation}
where $(\phi^+, \phi^0) \sim (1,2,1/2)$ is the usual Higgs scalar doublet.
The structure of this operator clearly shows that $any$ Majorana neutrino 
mass is $seesaw$ in character, i.e. of the form $v^2$ divided by an effective 
heavy mass, where $v$ is the vacuum expectation value (VEV) of $\phi^0$ as the 
electroweak gauge symmetry $SU(2)_L \times U(1)_Y$ is broken down to $U(1)_Q$. 
Different models of neutrino mass are merely different realizations \cite{ma} 
of this operator.

\section{Canonical Seesaw and Higgs Triplet Mechanisms for Neutrino Masses 
and Leptogenesis}

The most famous mechanism for getting a small $m_\nu$ is the canonical 
seesaw \cite{seesaw} where a heavy singlet neutral fermion $N$ is inserted 
between the two factors of Eq.~(1) with a large Majorana mass $m_N$.  Hence 
one may read off the neutrino mass as $m_\nu = f_i f_j v^2/m_N$.  An equally 
$simple$ and $natural$ mechanism \cite{triplet} is to realize Eq.~(1) with a 
heavy Higgs scalar triplet $(\xi^{++}, \xi^+, \xi^0)$ with couplings $f_{ij}$ 
to 2 lepton doublets and $\mu$ to 2 Higgs doublets.  The neutrino mass matrix 
is then given by $2 f_{ij} \mu v^2/m_\xi^2$.  This may be interpreted also 
as $\nu_i \nu_j$ coupling to the VEV of $\xi^0$ , which shows clearly 
the important point that it is possible as well as natural for $\langle \xi^0 
\rangle$ to be very much less than $m_\xi$.

Both of these two simple neutrino-mass mechanisms are also ideal for 
leptogenesis.  The heavy singlet neutral fermion $N$ may decay into 
$e^- \phi^+$ with lepton number $L = 1$ or $e^+ \phi^-$ with $L = -1$. 
With 2 or more $N$'s, the one-loop corrections (involving both vertex and 
self-energy graphs) allow for CP violation in their interference with the 
tree graph, and may generate \cite{fuya} a lepton asymmetry of the Universe 
if the decay of the lightest $N$ occurs out of thermal equilibrium as the 
Universe expands and cools.  The heavy $\xi^{++}$ may decay into $e^+ e^+$ 
with $L = -2$ or $\phi^+ \phi^+$ with $L = 0$.  Again, with 2 or more $\xi$'s, 
the one-loop (self-energy only) graph allows for CP violation and creates 
\cite{triplet} a lepton asymmetry, i.e.
\begin{equation}
|A + i B|^2 - |A^* + i B^*|^2 = 4 {\rm Im} (A B^*).
\end{equation}

\section{$R$ Parity Violating Supersymmetry and Neutrino Masses}

I now come to my main topic which is the generation of neutrino masses 
through $R$ parity violation in supersymmetry \cite{hasu}.  The well-known 
superfield content of the Minimal Supersymmetric Standard Model (MSSM) is 
given by
\begin{eqnarray}
&& Q_i = (u_i, d_i)_L \sim (3,2,1/6), ~~u^c_i \sim (3^*,1,-2/3), ~~d^c_i 
\sim (3^*,1,1/3), \\  && L_i = (\nu_i, l_i)_L \sim (1,2,-1/2), ~~l^c_i 
\sim (1,1,1); \\ && H_1 = (h_1^0, h_1^-) \sim (1,2,-1/2), ~~H_2 = 
(h_2^+, h_2^0) \sim (1,2,1/2).
\end{eqnarray}
Given the above transformations under the standard $SU(3) \times SU(2) \times 
U(1)$ gauge group, the corresponding superpotential should contain in general 
all gauge-invariant bilinear and trilinear combinations of the superfields. 
However, to forbid the violation of both baryon number $B$ and lepton 
number $L$, each particle is usually assigned a dicrete $R$ parity
\begin{equation}
R \equiv (-1)^{3B+L+2j},
\end{equation}
which is assumed to be conserved by the allowed interactions.  Hence the 
MSSM superpotential has only the terms $H_1 H_2$, $H_1 L_i l^c_j$, 
$H_1 Q_i d^c_j$, and $H_2 Q_i u^c_j$.  Since the superfield $\nu^c_i \sim 
(1,1,0)$ is absent, $m_\nu = 0$ in the MSSM as in the minimal Standard Model. 
Neutrino oscillations \cite{osc1,osc2,osc3} are thus unexplained.

Phenomenologically, it makes sense to require only $B$ conservation (to make 
sure that the proton is stable), but to allow $L$ violation (hence $R$ parity 
violation) so that the additional terms $L_i H_2$, $L_i L_j l^c_k$, and 
$L_i Q_j d^c_k$ may occur.  Note that they all have $\Delta L = 1$.  Neutrino 
masses are now possible \cite{numass} with Eq.~(1) realized in at least 
3 ways.  

The first way is to use the bilinear terms
\begin{equation}
-\mu H_1 H_2 + \epsilon_i L_i H_2,
\end{equation}
from which a $7 \times 7$ neutralino-neutrino mass matrix is obtained:
\begin{equation}
{\cal M_N} = \left[ \begin{array}{c@{\quad}c@{\quad}c@{\quad}c@{\quad}c} 
M_1 & 0 & -g_1 v_1 & g_1 v_2 & -g_1 u_i \\ 0 & M_2 & g_2 v_1 & -g_2 v_2 & 
g_2 u_i \\ -g_1 v_1 & g_2 v_1 & 0 & -\mu & 0 \\ g_1 v_2 & -g_2 v_2 
& -\mu & 0 & \epsilon_i \\ -g_1 u_i & g_2 u_i & 0 & \epsilon_i & 0 
\end{array} \right],
\end{equation}
where $v_{1,2} = \langle h^0_{1,2} \rangle /2$ and $u_i = \langle \tilde 
\nu_i \rangle /2$, with $i = e, \mu, \tau$.  Note first that both 
$\epsilon_i$ and $u_i$ are nonzero in general.  Note also that even 
if $u_i/\epsilon_i$ is not the same for all $i$, only one linear combination 
of the three neutrinos gets a tree-level mass.  In terms of the effective 
operator of Eq.~(1), this is a tree-level realization with $\nu_i$ mixing 
with $\tilde h_1^0$ (through $\epsilon_i/\mu$) which then connects with 
$\langle h_1^0 \rangle$ and a linear combination of the $SU(2)_Y$ and 
$U(1)_Y$ gauginos. The latter has a soft supersymmetry breaking Majorana 
mass and acts just like $N$ in generating a small $m_\nu$.  Specifically,
\begin{equation}
m_{\nu_i} = - {(c^2 M_1 + s^2 M_2) g_1^2 (v_1 \epsilon_i + \mu u_i)^2 \over 
s^2 M_1 M_2 \mu^2 - 2 g_1^2 v_1 v_2 \mu (c^2 M_1 + s^2 M_2)},
\end{equation}
where $s \equiv \sin \theta_W$ and $c \equiv \cos \theta_W$.

The second way is to use the trilinear terms, from which neutrino masses are 
obtained \cite{numass} as one-loop radiative corrections.  Note that these 
occur as the result of supersymmetry breaking and are also suppressed by 
$m_d^2$ or $m_l^2$.  A typical graph connects $\nu_i$ and $\nu_j$ through 
the intermediate states $(b, \tilde b^c)$ and $(b^c, \tilde b)$ which are 
linked by 2 $\langle h_1^0 \rangle$'s, as required by Eq.~(1).  Here
\begin{equation}
m_\nu \sim {3 \lambda'^2 \over 16 \pi^2} {A m_b^2 \over m^2_{\tilde b}}.
\end{equation}
For $m_\nu \sim 0.05$ eV, this implies $\lambda' > 10^{-4}$ for 
$m^2_{\tilde b}/A > 100$ GeV.

The third way is to recognize the fact that the sneutrino $\tilde \nu$ 
may have a ``Majorana'' mass term, i.e. $m^2 \tilde \nu \tilde \nu + h.c.$, 
in addition to the usual ``Dirac'' mass term, i.e. $M^2 \tilde \nu^* \tilde 
\nu$.  This leads inevitably \cite{hkk} to $m_\nu \neq 0$, but the effect 
occurs in two loops and is usually negligible.  An interesting exception is 
in the case of the specific leptogenesis scenario \cite{hms} to be discussed 
below.

\section{$R$ Parity Violating Supersymmetry and Leptogenesis}

As noted earlier, the $R$ parity violating interactions have $\Delta L 
= 1$.  Furthermore, the particles involved have masses at most equal to the 
supersymmetry breaking scale, i.e. a few TeV.  This means that their 
$L$ violation together with the $B + L$ violation by sphalerons \cite{krs} 
would erase any primordial $B$ or $L$ asymmetry of the Universe \cite{erase}. 
To avoid such a possibility, one may reduce the relevant Yukawa couplings 
to less than about $10^{-7}$, but a typical minimum value of $10^{-4}$ (see 
previous section) is 
required for realistic neutrino masses.  Hence the existence of the present 
baryon asymmetry of the Universe is unexplained if neutrino masses originate 
from these $\Delta L = 1$ interactions.  This is a generic problem of all 
models of radiative neutrino masses where the $L$ violation can be traced 
to interactions occuring at energies below $10^{13}$ GeV or so.

Once the notion of $R$ parity violation is introduced, there are many new 
terms to be added in the Lagrangian.  Some may be responsible for realistic 
neutrino masses and may even participate in the erasure of any primordial 
$B$ or $L$ asymmetry of the Universe, but others may be able to produce a 
lepton asymmetry \cite{active} on their own which then gets converted into 
the present observed baryon asymmetry of the Universe through the sphalerons.

Consider the usual $4 \times 4$ neutralino mass matrix in the $(\tilde B, 
\tilde W_3, \tilde h_1^0, \tilde h_2^0)$ basis:
\begin{equation}
{\cal M_N} = \left[ \begin{array} {c@{\quad}c@{\quad}c@{\quad}c} M_1 & 0 & 
-s m_3 & s m_4 \\ 0 & M_2 & c m_3 & -c m_4 \\ -s m_3 & c m_3 & 0 & -\mu \\ 
s m_4 & -c m_4 & -\mu & 0 \end{array} \right],
\end{equation}
where $m_3 = M_Z \cos \beta$, 
$m_4 = M_Z \sin \beta$, and $\tan \beta = v_2/v_1$.  The above assumes 
that $\epsilon_i$ and $u_i$ are negligible in Eq.~(8), which is 
a good approximation because neutrino masses are so small.  I now choose 
the special case of
\begin{equation}
m_3, ~m_4 << M_2 < M_1 < \mu.
\end{equation}
As a result, the two higgsinos $\tilde h^0_{1,2}$ form a heavy Dirac particle 
of mass $\mu$ and the other two less heavy Majorana fermion mass eigenstates 
are
\begin{eqnarray}
\tilde B' &\simeq& \tilde B + {sc \delta r_1 \over M_1-M_2} \tilde W_3 + ..., 
\\ \tilde W'_3 &\simeq& \tilde W_3 - {sc \delta r_2 \over M_1-M_2} \tilde B 
+ ...,
\end{eqnarray}
where $\delta = M_Z^2 \sin 2 \beta / \mu$, and
\begin{equation}
r_{1,2} = {1 + M_{1,2}/\mu \sin 2 \beta \over 1 - M_{1,2}^2/\mu^2}.
\end{equation}

I now observe that whereas $\tilde B$ couples to both $\bar l_L \tilde l_L$ 
and $\bar l^c_L \tilde l^c_L$, $\tilde W_3$ couples only to $\bar l_L \tilde 
l_L$ because $l^c_L$ is trivial under $SU(2)_L$.  On the other hand, 
$R$ parity violation implies that there is $\tilde l_L - h^-$ mixing as 
well as $\tilde l^c_L - h^+$ mixing.  Therefore, both $\tilde B'$ and 
$\tilde W'_3$ decay into $l^\pm h^\mp$ and may be the seeds of a lepton 
asymmetry in such a scenario.

Let the $\tilde l_L - h^-$ mixing be very small (which is a consistent 
assumption for realistic neutrino masses from bilinear $R$ parity violation). 
Then $\tilde W'_3$ decays only through its $\tilde B$ component.  Hence the 
decay rate of the LSP (Lightest Supersymmetric Particle), i.e. $\tilde W'_3$, 
is very much suppressed, first by $\delta$ and then by the $\tilde l^c_L - 
h^+$ mixing which will be denoted by $\xi$.  This construction is aimed at 
satisfying the out-of-equilibrium condition:
\begin{equation}
\Gamma (\tilde W'_3 \to l^\pm h^\mp) < H = 1.7 \sqrt {g_*} (T^2/M_{Pl})
\end{equation}
at the temperature $T \sim M_2$, where $H$ is the Hubble expansion rate of 
the Universe with $g_*$ the effective number of massless degrees of freedom 
and $M_{Pl}$ the Planck mass.  This implies
\begin{equation}
\left( {\xi |\delta| r_2 \over M_1-M_2} \right)^2 {1 \over M_2} < 1.9 \times 
10^{-14} {\rm GeV}^{-1},
\end{equation}
where $g_* = 10^2$ and $M_{Pl} = 10^{18}$ GeV.

The lepton asymmetry generated from the decay of $\tilde W'_3$ has both 
vertex and self-energy loop contributions from the insertion of $\tilde B'$. 
However, the coupling of $\tilde B'$ to $l^\pm h^\mp$ is suppressed only by 
$\xi$ and not by $\delta$, thus a realistic asymmetry may be established if 
$\xi$ is not too small.  Let $x \equiv M_2^2/M_1^2$, then the decay 
asymmetry of $\tilde W'_3$ is given by
\begin{equation}
\epsilon = {\alpha \xi^2 \over 2 \cos^2 \theta_W} {Im \delta^2 \over 
|\delta|^2} {\sqrt x g(x) \over 1-x},
\end{equation}
where
\begin{equation}
g(x) = 1 + {2(1-x) \over x}\left[ \left( {1+x \over x} \right) \ln (1+x) - 1 
\right],
\end{equation}
and Im$\delta$ comes from the relative phase between $M_1$ and $M_2$. 

At $T < M_2$, a lepton asymmetry may start to appear, but there are also 
reactions which destroy it: (I) recombination (inverse decay), i.e.
\begin{equation}
l^\pm + h^\mp \to \tilde W'_3 ~({\rm weak}), ~~~l^\pm + h^\mp \to \tilde B' 
~({\rm strong});
\end{equation}
(II) scattering, i.e. $l^\pm + h^\mp \to l^\mp + h^\pm$ with $\tilde W'_3$ 
(negligible) and $\tilde B'$ (weak) as intermediate states; and (III) 
annihilation, i.e. $\tilde W + \tilde W \to W + W$ which is $L$ and $R$ 
conserving (weak).  The Boltzmann equations must then be numerically solved 
to see if a lepton asymmetry ($\epsilon_L = n_B/g_* n_\gamma$) of order 
10$^{-10}$ can be generated for a given set of input parameters.  The choice 
of values for $M_1$ and $M_2$ is crucial for this purpose, because the 
inverse decay of $\tilde B'$ is capable of depleting $\epsilon_L$ by several 
orders of magnitude.  For example, if $M_1 = 3$ TeV and $M_2 = 2$ TeV, then 
$\epsilon_L \sim 10^{-14}$.

Two scenarios which work are \cite{hms}
\begin{eqnarray}
{\rm (A)} && M_2 = 3.5 ~{\rm TeV}, ~~M_1 = 6 ~{\rm TeV}, ~~\mu = 10 ~{\rm TeV},
\nonumber \\ && \xi = 5 \times 10^{-3}, ~~\sin 2 \beta = 0.10, ~~m_h = 200 
~{\rm GeV}; \\ {\rm (B)} && M_2 = 2 ~{\rm TeV}, ~~M_1 = 5 ~{\rm TeV}, ~~\mu = 
7.5 ~{\rm TeV}, \nonumber \\ && \xi = 5 \times 10^{-3}, ~~\sin 2 \beta = 0.05, 
~~m_h = 200 ~{\rm GeV}.
\end{eqnarray}

Hence realistic leptogenesis is possible if $\xi \sim 10^{-3}$ can be 
obtained.  This is actually not so easy because the origin of $\tilde l^c_L 
- h^+$ mixing in $R$ parity violation is usually the term $H_1 \tilde L 
\tilde l^c$, which is very small because $\langle \tilde \nu \rangle$ has 
to be very small.  To obtain $\xi \sim 10^{-3}$, it is necessary to add the 
nonholomorphic \cite{nonho} term $H_2^\dagger H_1 \tilde l^c$ which is 
generally unconstrained.  In the presence of this new term, the sneutrino 
$\tilde \nu$ also gets a ``Majorana'' mass of order 100 MeV, which then 
allows $m_\nu$ to be of order 10$^{-3}$ eV in 2 loops.

\section{Conclusion}

$\bullet$ In supersymmetry with $L$ violation (hence $R$ parity violation), 
realistic neutrino masses are obtained at 0, 1, and 2 loops.

\noindent $\bullet$ Successful leptogenesis is possible in a specific scenario:

(1) LSP is mostly $\tilde W_3$,

(2) gaugino masses $M_1$ and $M_2$ have a relative phase,

(3) $\tilde l_L - h^-$ mixing is negligible,

(4) $\tilde l_R - h^-$ mixing is ${\cal O}(10^{-3})$ from the 
nonholomorphic $H_2^\dagger H_1 \tilde l^c$ term.

\section*{Acknowledgments}
I thank Hans Klapdor and the other organizers of Dark 2000 for their great 
hospitality at Heidelberg.  This work was supported in part by the 
U.~S.~Department of Energy under Grant No.~DE-FG03-94ER40837.

\end{document}